\documentclass{npw}
\usepackage{graphicx}
\usepackage{bm}
\usepackage{citesort}

\begin{document}

\title{Multi-dimensional constraint relativistic mean field model
and applications in actinide and transfermium nuclei}
\author{
  Bing-Nan Lu$^{1}$, Jie Zhao$^{1}$, En-Guang Zhao$^{1,2}$ and 
  Shan-Gui Zhou$^{1,2}$\email{sgzhou@itp.ac.cn} \\
  \it $^{1}$State Key Laboratory of Theoretical Physics, Institute of Theoretical Physics, \\
  \it Chinese Academy of Sciences, Beijing 100190, China \\
  \it $^{2}$Center of Theoretical Nuclear Physics, National Laboratory of Heavy Ion Accelerator, \\
  \it Lanzhou 730000, China   
}
\date{Nov 14, 2013}
\maketitle

\begin{abstract}
In this contribution we present some results of potential energy surfaces of
actinide and transfermium nuclei from multi-dimensional constrained 
relativistic mean field (MDC-RMF) models.
Recently we developed multi-dimensional constrained covariant density functional 
theories (MDC-CDFT)
in which all shape degrees of freedom $\beta_{\lambda\mu}$ 
with even $\mu$ are allowed
and the functional can be one of the following four forms:
the meson exchange or point-coupling nucleon interactions combined with
the non-linear or density-dependent couplings.
In MDC-RMF models, the pairing correlations are treated with the BCS method.
With MDC-RMF models, the potential energy surfaces of even-even actinide nuclei
were investigated and the effect of triaxiality on the fission barriers 
in these nuclei was discussed.
The non-axial reflection-asymmetric $\beta_{32}$ shape in some transfermium
nuclei with $N=150$, namely $^{246}$Cm, $^{248}$Cf, $^{250}$Fm, and $^{252}$No
were also studied.
\end{abstract}

\section{Introduction}
\label{sec:intro}

\begin{figure}
\begin{center}
\resizebox{1.0\columnwidth}{!}{%
 \includegraphics{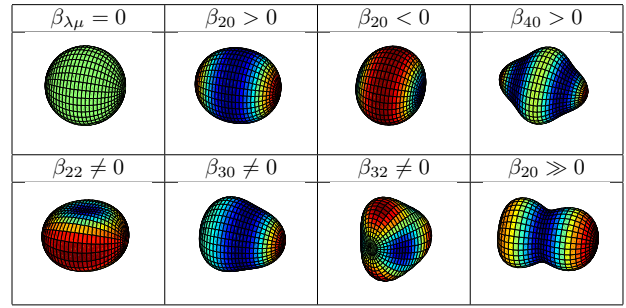} }
\end{center}
\caption{\label{Pic:deformations}(Color online)
A schematic show of some typical nuclear shapes. 
From left to right, the 1st row: (a) Sphere, (b) Prolate spheroid,
(c) Oblate spheroid, (d) Hexadecapole shape, 
and the second row: (e) Triaxial ellipsoid,
(f) Reflection symmetric octupole shape, (g) Tetrahedron,
(h) Reflection asymmetric octupole shape with very large quadrupole deformation
    and large hexadecapole deformation.
Taken from Ref.~\cite{Lu2012_PhD}.
}
\end{figure}

The occurrence of spontaneous symmetry breaking in atomic nuclei leads to 
various nuclear shapes
which can usually be described by the parametrization of the nuclear surface or the
nucleon density distribution~\cite{Bohr1998_Nucl_Structure_1,Ring1980}.
In mean-field calculations, the following parametrization
\begin{equation}
 \beta_{\lambda\mu} = {4\pi \over 3AR^\lambda} \langle Q_{\lambda\mu} \rangle,
 \label{eq:01}
\end{equation}
is usually used, where $Q_{\lambda\mu}$ are the mass multipole operators.
In Fig.~\ref{Pic:deformations}, a schematic show is given for some typical nuclear shapes.
The majority of observed nuclear shapes is of spheroidal form which
is usually described by $\beta_{20}$.
Higher-order deformations with $\lambda>2$ such as $\beta_{30}$
also appear in certain atomic mass regions~\cite{Butler1996_RMP68-349}.
In addition, non-axial shapes in atomic nuclei, in particular,
the nonaxial-quadrupole (triaxial) deformation $\beta_{22}$ has been
studied both experimentally and
theoretically~\cite{Starosta2001_PRL86-971,Odegard2001_PRL86-5866,Meng2010_JPG37-064025}.
The influence of the nonaxial octupole $\beta_{32}$ deformation
on the low-lying spectra has been also investigated~\cite{Hamamoto1991_ZPD21-163,Skalski1991_PRC43-140,%
Li1994_PRC49-R1250,Takami1998_PLB431-242,Yamagami2001_NPA693-579,%
Dudek2002_PRL88-252502,Dudek2006_PRL97-072501,Olbratowski2006_IJMPE15-333,%
Zberecki2006_PRC74-051302R,Dudek2010_JPG37-064032}.

In nuclear fission study, 
various shape degrees of freedom play important and different roles
in the occurrence and in determining the heights of the inner and outer barriers
in actinide nuclei (in these nuclei double-humped fission barriers usually appear).
For example, the inner fission barrier is usually lowered when the triaxial
deformation is allowed, while for the outer barrier the reflection asymmetric
(RA) shape is favored~\cite{Pashkevich1969_NPA133-400,Moeller1970_PLB31-283,%
Girod1983_PRC27-2317,Rutz1995_NPA590-680,Abusara2010_PRC82-044303,%
Prassa2012_PRC86-024317,Prassa2013_PRC88-044324}.

In order to give a microscopic and self-consistent description of
the potential energy surface (PES) with more shape degrees of freedom included,
multi-dimensional constrained covariant density functional theories were developed
recently~\cite{Lu2012_PRC85-011301R,Lu2013_arXiv1304.2513}.
In these theories, all shape degrees of freedom $\beta_{\lambda\mu}$ 
with even $\mu$ are allowed.
In this contribution, we present two recent applications of these theories:
the PES's of actinide nuclei and
the non-axial reflection-asymmetric $\beta_{32}$ shape in some transfermium
nuclei.
In Section~\ref{sec:formalism}, the formalism of our multi-dimensional constrained
covariant density functional theories will be given briefly.
The results and discussions are presented in Section~\ref{sec:results}.
Finally we give a summary in Section~\ref{sec:summary}.


\section{Formalism}
\label{sec:formalism}

The details of the formalism for covariant density functional theories
can be found in Refs.~\cite{Serot1986_ANP16-1,Reinhard1989_RPP52-439,%
Ring1996_PPNP37-193,Vretenar2005_PR409-101,Meng2006_PPNP57-470,Niksic2011_PPNP66-519}.
The CDFT functional in our multi-dimensional constrained calculations
can be one of the following four forms:
the meson exchange or point-coupling nucleon interactions combined with
the non-linear or density-dependent couplings~\cite{Lu2012_PRC85-011301R,Lu2013_arXiv1304.2513,%
Lu2012_EPJWoC38-05003,Lu2013_arXiv1304.6830}.
Here we show briefly the one corresponding to the non-linear point coupling
(NL-PC) interactions.
The starting point of the relativistic
NL-PC density functional is the following Lagrangian:
\begin{equation}
 \mathcal{L} = \bar{\psi}(i\gamma_{\mu}\partial^{\mu}-M)\psi
              -\mathcal{L}_{{\rm lin}}
              -\mathcal{L}_{{\rm nl}}
              -\mathcal{L}_{{\rm der}}
              -\mathcal{L}_{{\rm Cou}},
\end{equation}
where
\begin{eqnarray}
 \mathcal{L}_{{\rm lin}} & = & \frac{1}{2} \alpha_{S} \rho_{S}^{2}
                              +\frac{1}{2} \alpha_{V} \rho_{V}^{2}
                              +\frac{1}{2} \alpha_{TS} \vec{\rho}_{TS}^{2}
                              +\frac{1}{2} \alpha_{TV} \vec{\rho}_{TV}^{2} ,
 \nonumber \\
 \mathcal{L}_{{\rm nl}}  & = & \frac{1}{3} \beta_{S} \rho_{S}^{3}
                              +\frac{1}{4} \gamma_{S}\rho_{S}^{4}
                              +\frac{1}{4} \gamma_{V}[\rho_{V}^{2}]^{2} ,
 \nonumber \\
 \mathcal{L}_{{\rm der}} & = & \frac{1}{2} \delta_{S}[\partial_{\nu}\rho_{S}]^{2}
                              +\frac{1}{2} \delta_{V}[\partial_{\nu}\rho_{V}]^{2}
                              +\frac{1}{2} \delta_{TS}[\partial_{\nu}\vec{\rho}_{TS}]^{2}
 \nonumber \\
 &  & \mbox{}                 +\frac{1}{2} \delta_{TV}[\partial_{\nu}\vec{\rho}_{TV}]^{2} ,
 \nonumber \\
 \mathcal{L}_{{\rm Cou}} & = & \frac{1}{4} F^{\mu\nu} F_{\mu\nu}
                             +e\frac{1-\tau_{3}}{2} A_{0} \rho_{V} ,
\label{eq:lagrangian}
\end{eqnarray}
are the linear coupling, nonlinear coupling, derivative coupling,
and the Coulomb part, respectively.
$M$ is the nucleon mass, $\alpha_{S}$, $\alpha_{V}$, $\alpha_{TS}$,
$\alpha_{TV}$, $\beta_{S}$, $\gamma_{S}$, $\gamma_{V}$, $\delta_{S}$,
$\delta_{V}$, $\delta_{TS}$, and $\delta_{TV}$ are coupling constants
for different channels and $e$ is the electric charge.
$\rho_{S}$, $\vec{\rho}_{TS}$, $\rho_{V}$, and $\vec{\rho}_{TV}$ are the isoscalar density,
isovector density, time-like components of isoscalar current, and time-like components of
isovector current, respectively.
The densities and currents are defined as
\begin{eqnarray}
      \rho  _{S} = \bar{\psi}\psi , & \qquad &
 \vec{\rho}_{TS} = \bar{\psi}\vec{\tau}\psi ,
 \nonumber \\
      \rho_{V} = \bar{\psi} \gamma^{0} \psi , & \qquad &
 \vec{\rho}_{TV} = \bar{\psi} \vec{\tau} \gamma^{0} \psi.
 \label{eq:densities}
\end{eqnarray}
Starting from the above Lagrangian, using the Slater determinants as
trial wave functions and neglecting the Fock term as well as the contributions
to the densities and currents from the negative energy levels, one
can derive the equations of motion for the nucleons,
\begin{equation}
 \hat{h}\psi_{i} = \left(\bm{\alpha}\cdot\vec{p}+\beta(M+S(\vec{r}))+V(\vec{r})\right)\psi_{i}
                 = \epsilon_{i}\psi_{i}
 ,
\end{equation}
where the potentials $V(\bm{r})$ and $S(\bm{r})$ are calculated as
\begin{eqnarray}
S & = & \alpha_{S}\rho_{S}+\beta_{S}\rho_{S}^{2}+\gamma_{S}\rho_{S}^{3}+\delta_{S}\triangle\rho_{S}\nonumber \\
 &  & +\left(\alpha_{TS}\rho_{TS}+\delta_{TS}\triangle\rho_{TS}\right)\tau_{3}  ,\\
V & = & \alpha_{V}\rho_{V}+\gamma_{V}\rho_{V}^{3}+\delta_{V}\triangle\rho_{V}W\nonumber \\
 &  & +\left(\alpha_{TV}\rho_{TV}+\delta_{TV}\triangle\rho_{TV}\right)\tau_{3}  .
\end{eqnarray}

An axially deformed harmonic oscillator (ADHO) basis is adopted for solving the
Dirac equation~\cite{Lu2012_PRC85-011301R,Lu2013_arXiv1304.2513,Lu2011_PRC84-014328}.
The ADHO basis is defined as the eigen solutions of the Schrodinger
equation with an ADHO potential~\cite{Gambhir1990_APNY198-132,Ring1997_CPC105-77},
\begin{eqnarray}
 \left[ -\frac{\hbar^{2}}{2M} \nabla^{2} + V_{B}(z,\rho) \right] \Phi_{\alpha}(\bm{r}\sigma)
 & = &
 E_{\alpha} \Phi_{\alpha}(\bm{r}\sigma)
 ,
 \label{eq:BasSchrodinger-1}
\end{eqnarray}
where
\begin{equation}
 V_{B}(z,\rho) = \frac{1}{2} M ( \omega_{\rho}^{2}\rho^{2} + \omega_{z}^{2}z^{2} )
 ,
\end{equation}
is the axially deformed HO potential and $\omega_{z}$ and $\omega_{\rho}$
are the oscillator frequencies along and perpendicular to $z$ axis, respectively.
These basis states are also eigen functions of the $z$ component of the
angular momentum $j_{z}$ with eigen values $K=m_{l}+m_{s}$.
For any basis state $\Phi_{\alpha}(\bm{r}\sigma)$, the time reversal state
is defined as $\Phi_{\bar{\alpha}}(\bm{r}\sigma)=\mathcal{T}\Phi_{\alpha}(\bm{r}\sigma)$,
where $\mathcal{T}=i\sigma_{y}K$ is the time reversal operator and
$K$ is the complex conjugation.
Apparently we have $K_{\bar{\alpha}}=-K_{\alpha}$
and $\pi_{\bar{\alpha}}=\pi_{\alpha}$.
These basis states form a complete set for expanding any two-component spinors.
For a Dirac spinor with four components,
\begin{equation}
 \psi_{i}(\bm{r}\sigma) =
 \left( \begin{array}{c}
        \sum_{\alpha}f_{i}^{\alpha} \Phi_{\alpha}(\bm{r}\sigma) \\
        \sum_{\alpha}g_{i}^{\alpha} \Phi_{\alpha}(\bm{r}\sigma)
        \end{array}
 \right),
\end{equation}
where the sum runs over all the possible combination of the quantum
numbers $\alpha=\{n_{z},n_{r},m_{l},m_{s}\}$, and $f_{i}^{\alpha}$ and
$g_{i}^{\alpha}$ are the expansion coefficients.
In practical calculations, one should truncate the basis
in a proper way~\cite{Lu2012_PRC85-011301R,Lu2013_arXiv1304.2513,Lu2011_PRC84-014328}.

The nucleus is assumed to be symmetric under the $V_4$ group, that is, all the
potentials and densities can be expanded as 
\begin{equation}
 f(\rho,\varphi,z) =  f_{0}(\rho,z) \frac{1}{\sqrt{2\pi}}
+ \sum_{n=1}^{\infty} f_{n}(\rho,z) \frac{1}{\sqrt{\pi}}\cos(2n\varphi),
\end{equation}

The PES is obtained by the constrained self-consistent calculation,
\begin{equation}
 E^{\prime} = E_{{\rm RMF}} +
              \sum_{\lambda\mu} \frac{1}{2} C_{\lambda\mu}Q_{\lambda\mu} ,
\end{equation}
where the variables $C_{\lambda\mu}$'s change their values during the iteration.

Both the BCS approach and the Bogoliubov transformation are implemented
in our model to take into account the pairing effects.
For convenience, we name the MDC-CDFT with the BCS approach for the pairing 
as the MDC-RMF model and that with the Bogoliubov transformation as the MDC-RHB model.
More details of the multi-dimensional constraint covariant density
functional theories can be found in Refs.~\cite{Lu2012_PRC85-011301R,Lu2013_arXiv1304.2513}.


\section{Results and discussions}
\label{sec:results}

\subsection{PES's of actinides} 

\begin{figure}
\begin{center}
\resizebox{0.9\columnwidth}{!}{%
 \includegraphics{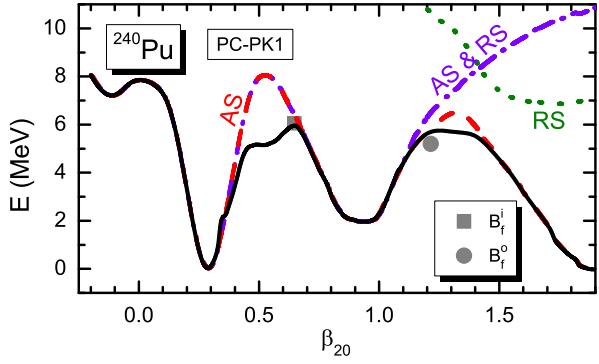} }
\end{center}
\caption{\label{Pic:PU240-1d}(Color online)
Potential energy curves of $^{240}$Pu with various self-consistent symmetries imposed.
The solid black curve represents the calculated fission path with $V_4$ symmetry imposed:
the red dashed curve is that with axial symmetry (AS) imposed,
the green dotted curve that with reflection symmetry (RS) imposed,
the violet dot-dashed line that with both symmetries (AS \& RS) imposed.
The empirical inner (outer) barrier height $B_\mathrm{emp}$ is denoted by the grey square (circle).
The energy is normalized with respect to the binding energy of the ground state.
The parameter set used is PC-PK1.
Taken from Ref.~\cite{Lu2012_PRC85-011301R}.
}
\end{figure}

In Refs.~\cite{Lu2012_PRC85-011301R,Lu2013_arXiv1304.2513}, one- (1-d),
two- (2-d), and three-dimensional (3-d) constrained calculations were 
performed for the actinide nucleus $^{240}$Pu.
The MDC-RMF model with the parameter set PC-PK1~\cite{Zhao2010_PRC82-054319} was used.
In Fig.~\ref{Pic:PU240-1d} we show the 1-d potential energy curves 
from an oblate shape with $\beta_{20}$ about $-0.2$ to the fission
configuration with $\beta_{20}$ beyond 2.0
which are obtained from calculations with different self-consistent symmetries
imposed: the axial (AS) or triaxial (TS) symmetries combined with
reflection symmetric (RS) or asymmetric cases.
The importance of the triaxial deformation on the inner barrier and
that of the octupole deformation on the outer barrier are clearly seen:
The triaxial deformation reduces the inner barrier height by more than 2 MeV
and results in a better agreement with the empirical value~\cite{Abusara2010_PRC82-044303};
the RA shape is favored beyond the fission isomer and lowers very much
the outer fission barrier~\cite{Rutz1995_NPA590-680}.
Besides these features, it was found for the first time that the outer
barrier is also considerably lowered by about 1 MeV when the triaxial
deformation is allowed.
In addition, a better reproduction of the empirical barrier height can be seen for
the outer barrier.
It has been stressed that this feature can only be found when the axial and
reflection symmetries are simultaneously broken~\cite{Lu2012_PRC85-011301R}.

\begin{figure}
\begin{center}
\resizebox{0.95\columnwidth}{!}{%
 \includegraphics{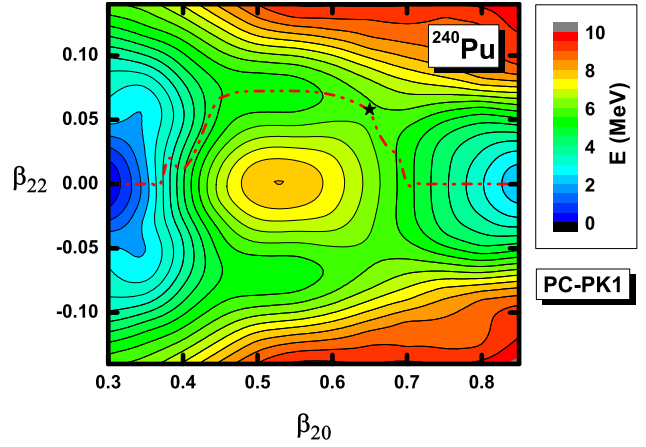} }
\end{center}
\caption{\label{Pic:PU240_2dI}(Color online)
Potential energy surfaces of $^{240}$Pu in the $(\beta_{20},\beta_{22})$ plane
around the inner barrier.
The energy is normalized with respect to the binding energy of the ground state.
The least-energy fission path is represented by a dash-dotted line.
The saddle point is denoted by the full star.
The contour interval is 0.5 MeV.
}
\end{figure}

\begin{figure}
\begin{center}
\resizebox{0.95\columnwidth}{!}{%
 \includegraphics{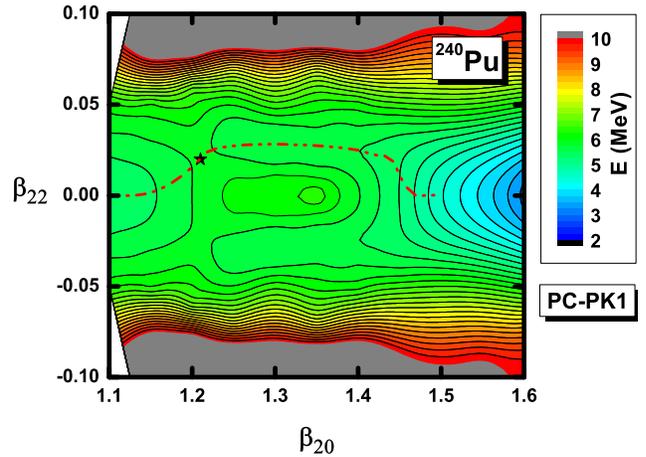} }
\end{center}
\caption{\label{Pic:PU240_2dO}(Color online)
Potential energy surfaces of $^{240}$Pu in the $(\beta_{20},\beta_{22})$ plane
around the outer barrier.
The energy is normalized with respect to the binding energy of the ground state.
The least-energy fission path is represented by a dash-dotted line.
The saddle point is denoted by the full star.
The contour interval is 0.25 MeV.
}
\end{figure}

Two-dimensional PES's in the $(\beta_{20},\beta_{22})$ plane near the inner and 
outer barriers are shown in Figs.~\ref{Pic:PU240_2dI} and~\ref{Pic:PU240_2dO}, respectively.
Starting from the axially symmetric ground state, the nucleus goes through the 
triaxial valley to the isometric state. 
The inner barrier is located at $\beta_{20} \approx 0.65$ and $\beta_{22} \approx 0.06$. 
The isomeric state keeps an axially symmetric shape.
As $\beta_{20}$ further increases, the nucleus goes through a triaxial valley
again, and then goes fission. 
The outer barrier is located at $\beta_{20} \approx 1.21$, $\beta_{22} \approx 0.02$, 
and $\beta_{30} \approx 0.37$.

A systematic study of even-even actinide nuclei has been carried out
and the results were presented in Ref.~\cite{Lu2013_arXiv1304.2513}
where we have shown that the triaxial deformation lowers
the outer barriers of these actinide nuclei by about $0.5 \sim 1$ MeV 
(about $10 \sim 20 \%$ of the barrier height).

\subsection{$Y_{32}$-correlations in $N=150$ isotones}

It has been anticipated that the tetrahedral shape 
($\beta_{\lambda\mu} = 0$, if $\lambda\ne3$ and $\mu\ne2$) may appear in the ground states of
some nuclei with special combinations of the neutron and
proton numbers~\cite{Li1994_PRC49-R1250,Dudek2010_JPG37-064032,Dudek2002_PRL88-252502}.
The tetrahedral symmetry-driven quantum effects may also lead to a large increase 
of binding energy in superheavy nuclei~\cite{Chen2013_NPR30-278}. 
However, no solid experimental evidence has been found for nuclei with
tetrahedral shapes.
On the other hand, $\beta_{32}$ deformation may appear together with other shape
degrees of freedome, say, $\beta_2$.
For example, it has been proposed that the non-axial
octupole $Y_{32}$-correlation results in the experimentally observed low-energy $2^-$
bands in the $N = 150$ isotones~\cite{Robinson2008_PRC78-034308} and the RASM calculations
reproduces well the experimental observables of these $2^-$ bands~\cite{Chen2008_PRC77-061305R}.

In Ref.~\cite{Zhao2012_PRC86-057304} the non-axial reflection-asymmetric $\beta_{32}$
deformation in $N=150$ isotones, namely $^{246}$Cm, $^{248}$Cf,
$^{250}$Fm, and $^{252}$No was investigated using the MDC-RMF model
with the parameter set DD-PC1~\cite{Niksic2008_PRC78-034318}.
It was found that 
due to the interaction between a pair of neutron orbitals,
$[734]9/2$ originating from $\nu j_{15/2}$ and
$[622]5/2$ originating from $\nu g_{9/2}$, and
that of a pair of proton orbitals,
$[521]3/2$ originating from $\pi f_{7/2}$ and
$[633]7/2$ originating from $\pi i_{13/2}$,
rather strong non-axial octupole $Y_{32}$ effects appear in $^{248}$Cf and
$^{250}$Fm which are both well deformed with large axial-quadrupole deformations,
$\beta_{20} \approx 0.3$.

\begin{figure}
\begin{center}
\resizebox{0.90\columnwidth}{!}{%
 \includegraphics{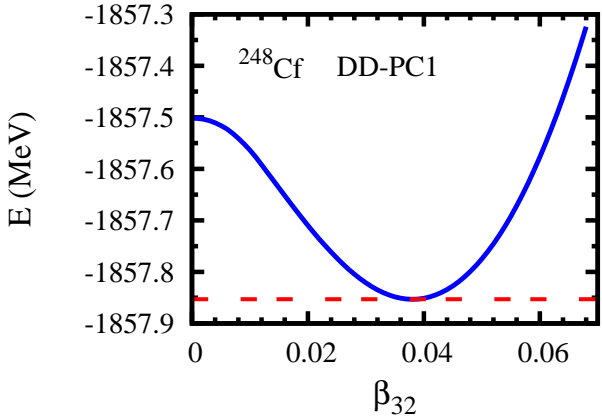} }
\end{center}
\caption{\label{fig:b32} (Color online)
The binding energy $E$ (relative to the ground state) for $^{248}$Cf
as a function of the non-axial octupole deformation parameter $\beta_{32}$.
}
\end{figure}

In Fig.~\ref{fig:b32}, the potential energy curve,
i.e., the total binding energy as a function of $\beta_{32}$ was shown
for $^{248}$Cf.
At each point of the potential energy curve, the energy is minimized
automatically with respect to other shape degrees
of freedom such as $\beta_{20}$, $\beta_{22}$, $\beta_{30}$, and $\beta_{40}$, etc.
One finds in this curve a clear pocket with the depth more than 0.3 MeV.
Similar potential energy curve was also obtained for $^{250}$Fm.
For $^{246}$Cm and $^{252}$No, only a shallow minimum develops along
the $\beta_{32}$ shape degree of freedom.
It was also shown that the evolution of the non-axial octupole $\beta_{32}$ effect along
the $N=150$ isotonic chain is not very sensitive to the form of
the energy density functional and the parameter set we used~\cite{Zhao2012_PRC86-057304}.

Both the non-axial octupole parameter $\beta_{32}$ and the energy gain due to
the $\beta_{32}$-distortion reach maximal values at $^{248}$Cf 
in the four nuclei along the $N=150$ isotonic chain~\cite{Zhao2012_PRC86-057304}.
This is consistent with the analysis given in
Refs.~\cite{Chen2008_PRC77-061305R,Jolos2011_JPG38-115103} and the experimental observation
that in $^{248}$Cf, the $2^-$ state is the lowest among these
nuclei~\cite{Robinson2008_PRC78-034308}.
These results indicate a strong $Y_{32}$-correlation in these nuclei.


\section{Summary}
\label{sec:summary}

In this contribution we present the formalism and some applications of the multi-dimensional
constrained relativistic mean field (MDC-RMF) models in which all shape degrees of freedom
$\beta_{\lambda\mu}$ with even $\mu$ are allowed.
The potential energy surfaces (curves) of actinide nuclei 
and the effect of the triaxiality on the first and second fission barriers 
were investigated.
It is found that besides the octupole deformation, the triaxiality also
plays an important role upon the second fission barriers.
The non-axial reflection-asymmetric $\beta_{32}$ shape in $N=150$ isotones were studied
and rather strong non-axial octupole $Y_{32}$ effects have been found in $^{248}$Cf and
$^{250}$Fm which are both well deformed with large axial-quadrupole deformations,
$\beta_{20} \approx 0.3$.


\begin{ack}
This work has been supported by Major State Basic Research Development 
Program of China (Grant No. 2013CB834400), 
National Natural Science Foundation of China (Grant Nos. 11121403, 11175252, 
11120101005, 11211120152, and 11275248), 
the Knowledge Innovation Project of Chinese Academy of Sciences (Grant No. KJCX2-EW-N01). 
The results described in this paper are obtained on the ScGrid of Supercomputing
Center, Computer Network Information Center of Chinese Academy of Sciences.
\end{ack}




\providecommand{\newblock}{}

\end{document}